\documentstyle[psfig,11pt,a4]{article}

\newcommand{\be}{\begin{equation}}
\newcommand{\ee}{\end{equation}} 
\newcommand{\bea}{\begin{eqnarray}}
\newcommand{\eea}{\end{eqnarray}}

\catcode`\@=11 
\@addtoreset{equation}{section}
 
\catcode`\@=11

\begin{document}

\begin{titlepage}

\begin{flushright} 
{\tt 	FTUV/96-41\\ 
	IFIC/96-49\\ 
	hep-th/9607155}
 \end{flushright}

\bigskip

\begin{center}

{\bf{\LARGE Black Hole Evaporation \\
by Thermal Bath Removal
\footnote{Work partially supported by the 
{\it Comisi\'on Interministerial de Ciencia y Tecnolog\'{\i}a}\/ 
and {\it DGICYT}.}}
}

\bigskip 
\bigskip\bigskip
 J.Cruz\footnote{\sc cruz@lie.uv.es} and
 J.Navarro-Salas\footnote{\sc jnavarro@lie.uv.es}

\end{center}

\bigskip%

\footnotesize
	
 Departamento de F\'{\i}sica Te\'orica and 
	IFIC, Centro Mixto Universidad de Valencia-CSIC.
	Facultad de F\'{\i}sica, Universidad de Valencia,	
        Burjassot-46100, Valencia, Spain.

\normalsize 

\bigskip

\bigskip

\begin{center}
{\bf Abstract}
\end{center}

We study the evaporation process of 2D black holes in thermal equilibrium when the 
incoming radiation is turned off.
Our analysis is based on two different classes of 2D dilaton gravity models which are exactly solvable 
in the semiclassical aproximation including back-reaction. We consider a one parameter family of 
models interpolating between the Russo-Susskind-Thorlacius and Bose-Parker-Peleg models.
We find that the end-state geometry is the same as the one coming from an evaporating black hole 
formed by gravitational collapse.  
We also study the quantum evolution
 of black holes arising in a model with classical action
$S={1\over2\pi}\int d^2x\sqrt{-g}\left(R\phi+4\lambda^2e^{\beta\phi}\right)$.
The black hole temperature is proportional to the mass and the exact semiclassical 
solution indicates that these black holes never disappear completely.



\end{titlepage}

\newpage

\section{Introduction}
 One of the main features of 1+1 dilaton gravity \cite{CGHS}
 (see the reviews \cite{Strominger} \cite{Gid} \cite{Thorlacius}) is 
 that the formation and semiclassical evaporation
 of a black hole can be studied in an exact analytical setting.
 The Russo-Susskind-Thorlacius (RST) model \cite{RST} predicts, with
  a natural choice of boundary conditions,
 an evaporation process ending in the linear dilaton vacuum in a finite proper time.
 Recently, Bose-Parker-Peleg \cite{BPP} have proposed a 
 different one-loop quantum modification of the CGHS model
 \cite{CGHS} describing a semiclassical evaporation with a 
 semi-infinite throat as the remnant geometry.
 In both models the matching between the evaporating black hole 
 solution and the static end-state geometry
 produces a thunderpop carrying a small amount of negative energy.
 The same physical picture of black hole evaporation exhibits the one-parameter family of models 
 interpolating between    
 the RST and BPP models \cite{area}.
 In this paper we consider a different scenario to study the
  process of black hole evaporation in these theories.
 Instead of analyzing the formation and subsequent evaporation of a black hole we shall consider the
  dynamical evolution of a black hole in thermal equilibrium 
when the incoming flux decreases.
We shall find that the above conventional picture is maintained.
 The evaporation leads to the same remnant geometry as the process initiated with
the gravitational collapse although now the energy of the thunderpop turns out to be 
independent of the mass of the initial state.

We shall also study a similar process but within a different theory of 1+1 dilaton 
gravity. It was shown in Ref \cite{sym}
that the unique theories admitting special conformal symmetries are the models with a classical exponential potential
\begin{displaymath}
S={1\over2\pi}\int d^2x\sqrt{-g}\left(R\phi+4\lambda^2e^{2\beta\phi}\right)\>.
\end{displaymath}
The classical CGHS action can be recovered for $\beta=0$ through the field redefinitions $g_{\mu\nu}
=\bar g_{\mu\nu}e^{-2\bar\phi}$, $\phi=e^{-2\bar\phi}$.
In contrast with the CGHS theory, the black holes arising from the above theory $\left(\beta\neq 0\right)$
have a non-constant temperature proportional to the mass.
Due to the presence of a special symmetry $\left(g_{\mu\nu}\longrightarrow e^{-\beta\epsilon}
g_{\mu\nu}\ ,\ \phi\longrightarrow\phi+\epsilon\right)$,
it is possible to construct a semiclassical theory invariant under the above symmetry.
So, in this theory one can also find analytic solutions
 describing the evaporation of a black hole in thermal equilibrium
due to the decrease of the incoming thermal flux.

\section{Evaporation in the BPP-RST models}
The one-paramater family of models \cite{area} interpolating between
 the BPP and RST models is given by the
 action     
\be
S_0+S_P+{N\over24\pi}\int d^2x\sqrt{-g}\left[\left(1-2a\right)\left(\nabla\phi\right)^2+\left(a-1\right)\phi R\right]
\>,\label{ai}
\ee
where $S_0$ is the classical CGHS action, $S_P$ is the Polyakov term
 and $a$ is an arbitrary real paramater. For $a={1\over2}$ we recover the RST model and the BPP
 model requires $a=0$.
The equations of motion, in conformal gauge, are equivalent to
\be
\partial_+\partial_-\left(\rho-\phi\right)=0\>,\label{aii}
\ee
\be
\partial_+\partial_-\left(e^{-2\phi}+{Na\over12}\rho\right)+\lambda^2e^{2\left(\rho-\phi\right)}=0
\>\label{aiii}
\ee
\bea
&&e^{-2\phi}\left(4\partial_{\pm}\rho\partial_{\pm}\phi-2\partial_{\pm}^2\phi\right)
-{N\over12}\left(\left(\partial_{\pm}
\rho\right)^2-\partial_{\pm}^2\rho\right)\nonumber\\
&&+{N\over12}\left(\left(2a-1\right)\left(\partial_{\pm}\phi\right)^2+\left(a-1\right)\partial_{\pm}^2\phi
-2\left(a-1\right)\partial_{\pm}\phi\partial_{\pm}\rho\right)\nonumber\\
&&+T^f_{\pm\pm}-{N\over12}t_{\pm}=0\>,\label{aiv}
\eea
where $t_{\pm}\left(x^{\pm}\right)$ are the boundary terms arising from the Polyakov action.
In Kruskal gauge $\left(\rho=\phi\right)$, they reduce to
\be
\partial_+\partial_-\left(e^{-2\phi}+{Na\over12}\phi\right)+\lambda^2=0\>,\label{av}
\ee 
\be
\partial^2_{\pm}\left(e^{-2\phi}+{Na\over12}\phi\right)+T^f_{\pm\pm}
-{N\over12}t_{\pm}=0\>.\label{avi}
\ee
In the absence of matter $\left(T^f_{\pm\pm}=0\right)$, the general solution of (\ref{av}) (\ref{avi}) is easily found
to be
\bea
e^{-2\phi}+{Na\over12}\phi&=&-\lambda^2x^+x^-+{N\over12}\int^{x^+}\int^{x^+}t_{+}\left(x^+\right)
\nonumber\\&&+{N\over12}\int^{x^-}\int^{x^-}t_-\left(x^-\right)+bx^++cx^-+d\>,\label{avii}
\eea
with $b,c,d$ arbitrary constants.
Let us consider a solution with $b=c=t_{\pm}\left(x^{\pm}\right)=0\ ,\ d={M\over\lambda}$
\be
e^{-2\phi}+{Na\over12}\phi=-\lambda^2x^+x^-+{M\over\lambda}\>,\label{aviii}
\ee
corresponding to a black hole with a curvature singularity (we require
$\left.{M\over\lambda}-\alpha >0\right)$
\be
\alpha={Na\over24}\left(1-\log{Na\over24}\right)=-\lambda^2x^+x^-+{M\over\lambda}\>,\label{aix}
\ee
and an apparent horizon $\left(\partial_+\phi=0\right)$ located at
\be
x^-=0\>.\label{ax}
\ee
If we evaluate the flux of radiation
\be
<T^f_{\pm\pm}>=-{N\over12}\left(\left(\partial_{\pm}\rho\right)^2-\partial^2_{\pm}\rho+t_{\pm}\right)
\>,\label{axi}
\ee
in past and future null infinity using asymptotically flat coordinates
\newline
 $\left(
\lambda x^{\pm}=\pm e^{\pm\lambda\sigma^{\pm}}\right)$ we find -due to the anomalous transformation 
properties of $t_{\pm}
$- the constant thermal value
\be
<T^f_{\sigma^{\pm}\sigma^{\pm}}>={N\lambda^2\over48}\>.\label{axii}
\ee
So that (\ref{aviii}) represents a  radiating black hole in thermal equilibrium with
incoming radiation at temperature $T={\lambda\over2\pi}$ (i.e, the analogue of
the Hartle-Hawking state).

Now we shall try to find out a solution without incomig flux which can be continuously matched to
(\ref{aviii}) through the null line $x^+=x_0^+$.
We shall assume that the asymptotically flat coordinate $\sigma^+$ is
 the same as for the initial solution.
With this hypothesis we must take $t_{\sigma^+}=0$ as the no incoming radiation condition,
which implies $t_{x^+}={1\over4\left(x^+\right)^2}$ for $x^+>x^+_0$.
 Therefore, assuming now the boundary conditions
\be
t_{x^+}={1\over4\left(x^{+}\right)^2}\theta\left(x^+-x^+_0\right)\>,\label{axiii} 
\ee
\be
t_{x^-}=0\>,\label{axiv}
\ee
as the new input for (\ref{avii}) and demanding continuity to the metric and its derivative through $x^+=x^+_0$,
we obtain (for $x^+>x^+_0$) the dynamical solution $\left(k={N\over12}\right)$
\be
e^{-2\phi}+{Na\over12}\phi=-\lambda^2x^+\left(x^-+\Delta\right)-{k\over4}
\left(\log{x^+\over x_0^+}+1\right)+{M\over\lambda}
\>,\label{axv}
\ee
where $\Delta=-{k\over4\lambda^2x_0^+}$.

The new asymptotically flat coordinates $\left\{\tilde\sigma^+,\tilde\sigma^-\right\}$ are
\be
\lambda x^+=e^{\lambda\tilde\sigma^+}\>,\label{axvi}
\ee
\be
-\lambda\left(x^-+\Delta\right)=e^{-\lambda\tilde\sigma^-}\>.\label{axvii}
\ee
The equality of $\sigma^+$ and $\tilde\sigma^+$ confirms our previous assumption and thus 
guarantees that solution (\ref{axv}) has vanishing incomig radiation.

The curvature singularity of (\ref{axv}) is
\be
\alpha={k a\over2}\left(1-\log{k a\over2}\right)=-\lambda^2x^+\left(x^-+\Delta\right)-
{k\over4}\left(\log{x^+\over x^+_0}+1\right)+{M\over\lambda}\>,\label{axviii}
\ee
whereas the apparent horizon
\be
-\lambda^2x^+\left(x^-+\Delta\right)={k\over4}\>,\label{aixx}
\ee
recedes as the black hole evaporates and intersects the curvature singularity at the point
(see Fig.1)\be
x^+_{int}=x_0^+e^{{4\over k}\left({M\over\lambda}-\alpha\right)}\>,\label{axx}
\ee
\be
x^-_{int}={k\over4\lambda^2x_0^+}\left[1-e^{-{4\over k}\left({M\over\lambda}-\alpha\right)}\right]
\>.\label{axxi}
\ee
We notice that the evaporation time differs from that obtained in
the process originated by gravitational collapse  \cite{area}.

Following the conventional reasoning \cite{RST} \cite{BPP} we shall
 try to implement the cosmic censorship conjeture
by matching the evaporating black hole solution (\ref{axv}) with a stable
 (not-radiating) solution on the
null line
$x^-=x_{int}^-$.
>From \cite{area} we know that the stable solutions of (\ref{av}) (\ref{avi}) are
\be
e^{-2\phi}+{Na\over12}\phi=-\lambda^2x^+\left(x^-+\Delta\right)-{k\over4}\log\left(
-\lambda^2x^+\left(x^-+\Delta\right)\right)
+C\>,\label{axxii}
\ee
corresponding to the boundary conditions $t_{x^{\pm}}={1\over4\left(x^{\pm}\right)^2}$. Imposing
continuity to the metric through $x^-=x_{int}^-$ we obtain the value of the constant $C$
\be
C={k\over4}\left(\log{k\over4}-1\right)+\alpha\>.\label{axxiii}
\ee
The remnant geometry (\ref{axxii}) coincides exactly with the final state resulting from the evaporation of a black hole formed 
by collapse of a shock wave, as discussed in \cite{area}.
However, in the present process the energy of the thunderpop emanating from the intersecting 
point is independent of the initial state and takes the limiting value
$E_{thunderpop}=-{\lambda k\over4}$.

\section{Evaporation in the exponential models}

In this section we shall investigate the evaporation process of black holes within a different theory
but with boundary conditions similar to that considered in the previous section.
Our starting point is the semiclassical action of the exponential models
\cite{sym}
\bea
S&=&{1\over2\pi}\int d^2x\sqrt{-g}\left(\phi R+4\lambda^2e^{\beta\phi}-{1\over2}  
\sum_{i=1}^N\left(\nabla f_i\right)^2\right)\nonumber\\
&&+S_P+{N\beta\over 96\pi}\int d^2x\sqrt{-g}\left(\phi R+\beta\left(\nabla\phi\right)^2\right)\>,\label{bi}
\eea
where $S_P$ is the Polyakov action.
The local counterterm of (\ref{bi}) can be obtained from
 the Polyakov effective action 
respect to the metric invariant under the special conformal symmetry.
This way one ensures the invariance of the semiclassical action and so the existence of a free field.
The equations of motion of
(\ref{bi}), in conformal gauge, are
\be
\partial_+\partial_-\left(\rho-{\beta\over2}\phi\right)=0\>,\label{bii}
\ee
\be
\partial_+\partial_-\left[\left(1+{N\beta\over24}\right)\phi+{N\rho\over12}
\right]+\lambda^2e^{\left(2\rho+\beta\phi\right)}=0\>,\label{biii}
\ee
\bea
-\left(1+{N\beta\over12}\right)\partial_{\pm}^2\phi+2\left(1+{N\beta\over12}\right)
\partial_{\pm}\phi\partial_{\pm}\rho&&\nonumber\\+
{N\over12}\left[\left(\partial_{\pm}\rho\right)^2-
\partial_{\pm}^2\rho+
{\beta^2\over4}\left(\partial_{\pm}\phi\right)^2+t_{\pm}\left(x^{\pm}\right)\right]&=&T^f_{\pm\pm}
\>.\label{biv}
\eea
In Kruskal gauge $\left(2\rho=\beta\phi\right)$ these equations can be rewritten as
\be
\partial_+\partial_-2\rho=-\lambda^2\beta\gamma e^{4\rho}\>,\label{bv}
\ee
\be
e^{2\rho}\partial_{\pm}^2e^{-2\rho}=\beta\gamma\left(T_{\pm\pm}^f-{N\over12}t_{\pm}\right)\label{bvi}
\ee
where $\gamma={1\over1+{N\beta\over12}}$.

We see that $2\rho$ satisfies the Liouville equation (\ref{bv}).
Its general solution can be expressed as
\be
\rho={1\over4}\log {\partial_+A\partial_-B\over\left(1+\lambda^2\beta\gamma AB\right)^2}\>,\label{bvii}
\ee
with $A\left(B\right)$ an arbitrary function of the  $x^+\left(x^-\right)$ coordinate.
Inserting (\ref{bvii}) into the constrained equations (\ref{bvi}) we obtain Ricatti differential equations
with respect to $\log\partial_+A$ and $\log\partial_-B$.
\be
{1\over4}\left(\partial_+\log\partial_+A\right)^2-{1\over2}\partial_+^2\log\partial_+A=
\beta\gamma\left(T^f_{++}-{N\over12}t_{+}\right)\>,\label{bviii}
\ee
\be
 {1\over4}\left(\partial_-\log\partial_-B\right)^2-{1\over2}\partial_-^2\log\partial_-B=
 \beta\gamma\left(T^f_{--}-{N\over12}t_-\right)\>.\label{bix}
 \ee
Taking $T_{\pm\pm}^f=t_{\pm}=0$, one obtains the solution
 \be
 ds^2={-dx^+dx^-\over{\lambda^2\beta\gamma\over C}+Cx^+x^-}\>.
 \label{bx}
 \ee
 For $C<0$ (and $ \lambda^2\beta<0$) \cite{sym} the above metric represents
  a black hole with a singularity curve
 \be
 {\lambda^2\beta\gamma\over C}+Cx^+x^-=0\>,\label{bxi}
 \ee
 and an apparent horizon located at
 \be
 x^-=0\>.\label{bxii}
 \ee
 Proceeding in a similar way as in the past section we can calculate the 
 flux of ingoing and outgoing radiation in the asymptotically flat
 coordinates $\left\{\sigma^{\pm}\right\}$
 \be
 \pm\sqrt{|C|}x^{\pm}=e^{\pm\sqrt{|C|}\sigma^{\pm}}\>,\label{bxiii}
 \ee
 getting 
\be
<T^f_{\sigma^{\pm}\sigma^{\pm}}>={N|C|\over48}\>,\label{bxiv}
\ee
which corresponds to a radiating black hole in thermal equilibrium
with ingoing radiation at temperature $T={\sqrt{|C|}\over2\pi}$.
The black hole mass can be calculated immediately \cite{sym}: $M={2\over\beta\pi}\sqrt{|C|}$.
 Therefore, the Hawking temperature turns out to be proportional to the mass:
$T={\beta\over4}M$.
The existence of 2D black holes whose mass is proportional to their
temperature was noted in Ref.~\cite{MST}.
At this point, the next step would be to solve equations
(\ref{bviii}) and (\ref{bix}) with the modified boundary conditions (\ref{axiii}) (\ref{axiv})
and impose continuity to the resulting metric and its derivative through $x^+=x^+_0$.
However this leads to an excesively complicated analytical treatment.
In order to  
simplify the analysis we shall admit that,
in addition to the discontinuity in the boundary function $t_{x^+}$, there is
an incomig shock-wave at $x^+=x^+_0$
\be
T^f_{++}={1\over2x^+_0\beta\gamma}\delta\left(x^+-x^+_0\right)\>.\label{bxv}
\ee
With this assumption, and in the limit $N\beta>>1$,
 the solutions of (\ref{bviii}) and (\ref{bix}), for
 $x^+>x^+_0$, lead to
the metric
\be
ds^2=-{\left({x_0^+\over x^+}\right)^{1\over2}\over{\lambda^2\beta\gamma\over C}+Cx^+_0x^-
\left(\log{x^+\over x^+_0}+1\right)}dx^+dx^-\>.\label{bxvi}
\ee
The most relevant feature of this solution is that the singularity curve
\be
{\lambda^2\beta\gamma\over C}+Cx^+_0x^-\left(\log{x^+\over x_0^+}+1\right)=0\>,\label{bxvii}
\ee
and the horizon
\be
{\lambda^2\beta\gamma\over C}+Cx^+_0x^-\left(\log{x^+\over x_0^+}+1\right)+2Cx_0^+x^-=0\>,\label{bxviii}
\ee
do not intersect each other at any finite point (see Fig.2). Moreover,
  in contrast with the BPP-RST models the relation of the asymptotically
 flat coordinate $\tilde\sigma^+$ of (\ref{bxvi})
 and the Kruskal one
 $x^+$, which regulates the incomig flux, is not the same as for the
  initial solution (\ref{bx}).
 In fact, in past null infinity, this relation becomes 
 \be
 {dx^+\over\sqrt{|C|}\left(x_0^+x^+\right)^{{1\over2}}\left(\log{x^+\over x^+_0}+1\right)}=d\tilde\sigma^+
 \>.\label{bixx}
 \ee
  This tell us that the thermal bath has not been actually removed.
  Using (\ref{bixx}) we can calculate the value of $<T^f_{\tilde\sigma^+\tilde\sigma^+}>$
   in past null infinity for
  $x^+>x_0^+$
 \bea
  <T^f_{\tilde\sigma^+\tilde\sigma^+}>&=&-{N\over12}\left[{|C|x_0^+\over4\left(x^+\right)}
  \left(\log{x^+\over x^+_0}+1\right)^2- 
{1\over8\left(x^+\right)^2}\right.\nonumber\\
&&\left.-{3\over16\left(x^+\right)^2}\left({\log{x^+\over x_0^+}+3\over
\log{x^+\over x_0^+}+1}\right)^2\right]\>.\label{bxx}
\eea
We see that the flux goes to zero at late times $\left(x^+\longrightarrow\infty\right)$
whereas at $x^+=x^+_0$ 
\be
<T^f_{\tilde\sigma^+\tilde\sigma^+}>|_{x_0^+}=-{N\over12}\left[{|C|\over4}
-{29\over16\left(x^+_0\right)^2}\right]\>.\label{bxxi}
\ee
At this point it seems natural to choice the point $x_0^+$ which makes
(\ref{bxxi}) to coincide with the constant value of $<T^f_{\sigma^+\sigma^+}>={N\over48}|C|$
 of the
 initial solution (i.e, $x_0^+=\left.\sqrt{{29\over8|C|}}\right)$.
In this way the external flux starts to decrease at $x^+=x^+_0$
(see Fig.3), vanishes at retarded time $x_1^+$ and for $x^+>x_1^+$
becomes a small negative energy flux vanishing again at infinity.
 
In conclusion, the semiclassical solution implies that the black hole
evaporation never ceases. One can argue that the non-vanishing incoming flux   
for $x^+>x_0^+$ prevents the complete evaporation\footnote{To obtain an 
evaporating solution with a vanishing incoming flux an adequate ansatz
for $t_{\pm}$ is required.}. 
However after $x^+>x^+_1$ the flux becomes negative and produces the opposite effect and,
despite of this, the apparent horizon never meets the singularity.
This qualitative behaviour of the solution can be understood in terms
of thermodynamic arguments.
In the evaporation process the temperature slows down because it is proportional
to the mass and the third law of thermodynamics \cite{Wald} does not permit
to achieve the absolute zero temperature (i.e, the complete evaporation).

\bigskip

After completing this work we were informed that the black hole solution 
(\ref{bx}) also appears in Ref.~\cite{Mann}.

\bigskip

The authors wish to thank  M. Navarro and C. F. Talavera
for useful discussions.

\newpage
\centerline{\psfig{figure=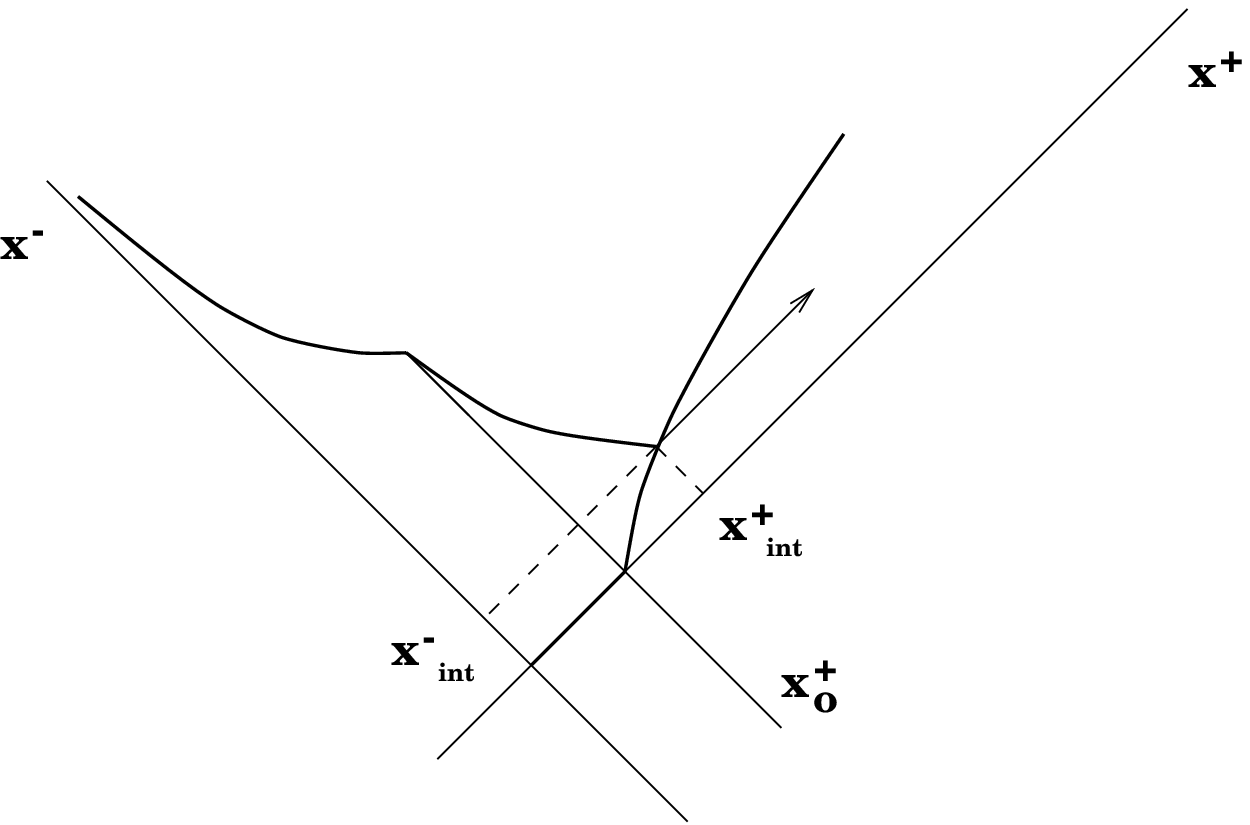,width=15cm}}
{\bf Fig.1}
Kruskal diagram of an evaporating black hole in the BPP-RST models
\centerline{\psfig{figure=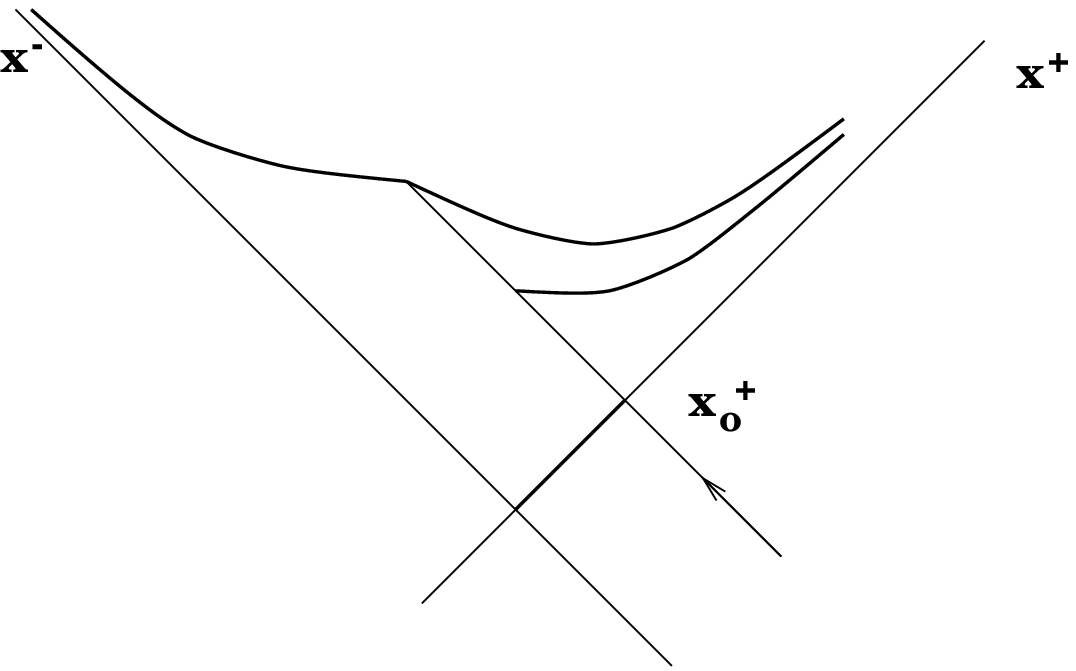,width=15cm}}
{\bf Fig.2} Kruskal diagram of
 an evaporating black hole in the exponential model
\centerline{\psfig{figure=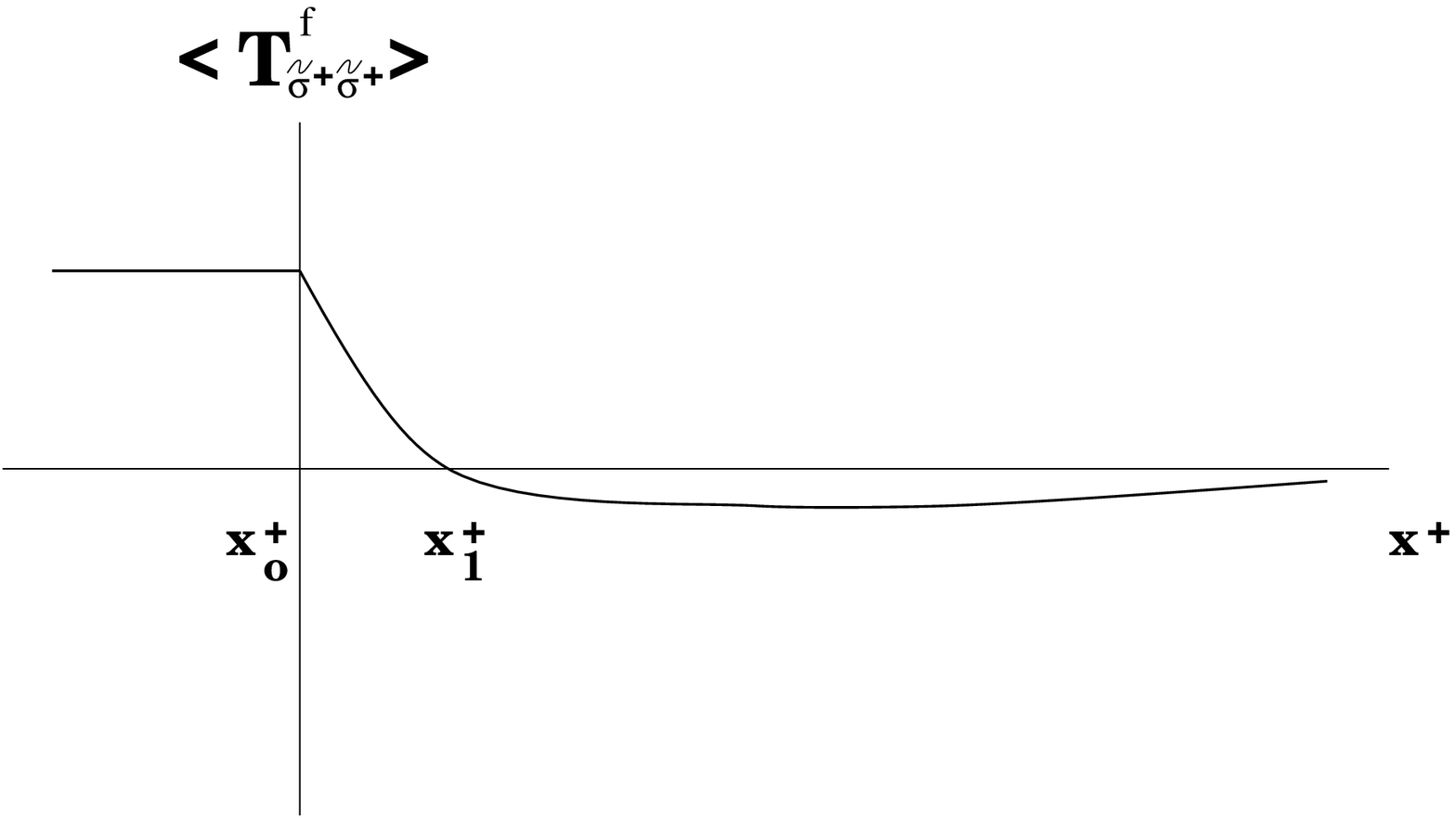,width=15cm}}
{\bf Fig.3}
Flux of incoming radiation associated with the boundary condition
$t_{x^+}={1\over4\left(x^+\right)^2}
\theta\left(x^+-x^+_0\right)$ in the exponential model
\end{document}